%% file: main.tex
\documentclass[twocolumn]{article}
\usepackage[utf8]{inputenc}
\usepackage{graphicx}
\usepackage{siunitx}
\usepackage{authblk}
\usepackage{hyperref}
\hypersetup{
    colorlinks=true,
    linkcolor=blue,
    filecolor=magenta,      
    urlcolor=cyan,
    }

\usepackage{geometry}
\title{\vspace{0mm} Droplet impact of blood and blood simulants on a solid surface: Effect of the deformability of red blood cells and the elasticity of plasma}

\author[1]{\small{Yuto Yokoyama}}
\author[2]{Akane Tanaka}
\author[1]{Yoshiyuki Tagawa}

\affil[1]{Department of Mechanical Systems Engineering, Tokyo University of Agriculture and Technology, Koganei Campus 6-204, 2-24-16 Nakacho, Koganei, Tokyo, Japan
}
\affil[2]{Laboratory of Comparative Animal Medicine, Division of Animal Life Science, Tokyo University of Agriculture and Technology, Fuchu Campus, 3-5-8 Saiwaicho, Fuchu, Tokyo, Japan
}

\date{}

\begin{document}

\twocolumn[

\maketitle

\vspace{-10mm}

\paragraph{Abstract} The impact of blood droplets onto a solid wall is of great importance for bloodstain pattern analysis in forensic science.
Previous studies suggest that the behaviour of impacting blood is similar to that of a Newtonian fluid, which has a shear viscosity equivalent to that of blood at high shear rates.
To understand this important fact, we conducted comparative experiments of droplet impact on a glass surface using whole blood and three solutions with a shear viscosity similar to that of blood. Specifically, we used dog's whole blood (deformable red blood cells dispersed in plasma, WB), plasma with non-deformable resin particles (PwP), glycerol and water with resin particles (GWwP), and a commercial blood simulant (hard particles dispersed in a water-based Newtonian solution, BS).
Ranges of Reynolds and Weber numbers in our experiments were 550 $<Re<$ 1700 and 120 $<We<$ 860, respectively.
Side and bottom views of droplet impact were simultaneously recorded by two high-speed cameras.
The spreading radius of the impacting WB droplet in our experiments agreed well with that of Newtonian fluids with viscosity similar to that of WB at high shear rates.
Splashing droplets of WB and Newtonian fluids form finger structures (finger-splashing).
Although PwP has a viscosity similar to that of WB at high shear rates, an impacting PwP droplet exhibited typical characteristics of impacting suspension droplets, that is, a reduced spreading radius and splashing with ejection of particles.
Such significant differences between impacting droplets of PwP and WB indicates that the high deformability of red blood cells in WB plays a crucial role in the Newtonian-like behaviour of blood droplets on impact.
The finger-splashing of PwP and GWwP exhibited no significant difference, indicating that the effect of plasma elasticity on finger-splashing is negligible.
Importantly, the impacting BS droplet behaved quite differently from WB in both spreading and splashing.
Our results imply that the use of deformable particles rather than hard particles in a BS is essential for mimicking blood droplet impact.

\vspace{20pt}

]

\input{text}

\section*{Acknowledgement}

This work was supported by JSPS KAKENHI Grant Numbers 20H00223, 20H00222, and 20K20972.
We thank R. Muko for helping with the preparation of the dog's blood and J. M. Gordillo and G. Riboux for fruitful discussion.

\bibliographystyle{ieeetr}
\bibliography{ref}

\end{document}

%% file: text.tex
\section{Introduction}\label{sec:intro}

\begin{figure}[t]
    \centering
    \includegraphics[width=0.95\columnwidth]{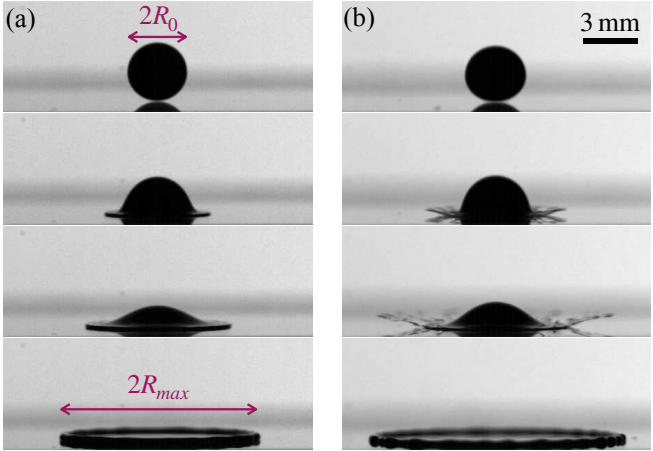}
    \caption{Blood droplets impacting onto a glass plate. (a) The deposited droplet spreads to maximum radius $R_{max}$ ($R_0=$ 1.6 \si{mm}, $V=$ 2.3 \si{m/s}). (b) Droplet splashes ($R_0=$ 1.6 \si{mm}, $V=$ 4.3 \si{m/s}).}
    \label{fig:droplet_impact}
\end{figure}

Bloodstains provide very important evidence to the investigation of violent crimes.
Bloodstain pattern analysis can be performed to estimate the location of the victim and the events that caused the bloodstain to appear, and to reconstruct the details of the crime scene \cite{adam2012, bevel2002}.When investigators reconstruct details of a crime scene from bloodstains, it is useful to know the impact velocity of the blood droplets that caused the stain \cite{hulse-smith2005}.
The impact velocity can be estimated from the size of the bloodstain caused by spreading (Fig. \ref{fig:droplet_impact}(a)) and the presence of satellite bloodstains around the main stain caused by splashing (Fig. \ref{fig:droplet_impact}(b)) \cite{hulse-smith2005,laan2014a,degoede2018,attinger2013}.
Since a bloodstain pattern is determined by the spreading and splashing behaviour during the droplet impact, it is essential to understand the impact behaviour of a blood droplet in detail.

Many theoretical models for Newtonian fluids have been developed to predict the ratio of the maximum radius $R_{max}$ to the initial radius of the droplet $R_0$ (spreading factor) \cite{clanet2004,pasandideh-fard1996,lee2016,yarin2006,visser2015} and the conditions under which splashing occurs (splashing threshold) \cite{mundo1995,bird2009,gordillo2019,josserand2016,hatakenaka2019,usawa2021}.
Previous studies have reported two types of splashing: corona splashing and prompt splashing \cite{rioboo2001,xu2007}.
Corona splashing is characterised by the lamella (liquid sheet that is ejected from the droplet) lifting off the solid surface, forming finger-like structures, and ejecting secondary droplets from the tips of the lamella.
Prompt splashing is characterised by ejection of fast and small droplets from the tip of the lamella which does not lift off the solid surface.
Corona splashing occurs to relatively viscous liquids with an Ohnesorge number above 0.0062 ($Oh = \eta/\sqrt{\rho R_0 \sigma}$, where $\rho$, $\eta$, and $\sigma$ are the density, viscosity, and surface tension of the liquid, respectively), while prompt splashing occurs with $Oh \leq 0.0062$ \cite{palacios2013,roisman2015,burzynski2020}.
In recent years, impacting droplets of non-Newtonian fluids, such as suspensions and viscoelastic fluids, have also been studied.
Grishaev {\it et al}. suggested that droplets of a suspension of resin particles have a reduced spreading factor as compared to Newtonian fluids with a similar viscosity due to the friction among the particles or between the particles and the solid surface \cite{grishaev2015}.
Peters {\it et al}. reported that during the splashing of a suspension droplet, dispersed ceramic particles are ejected (particle ejection) due to momentum transfer among particles \cite{peters2013a}.
Studies of elastic polymer solutions suggested that elasticity does not affect the spreading factor \cite{an2012}, but increases the threshold of prompt splashing \cite{vega2017}.
Blood is a dense suspension of mainly red blood cells in plasma \cite{sousa2016}.
However, previous studies have reported that the spreading factor $R_{max}/R_0$ of blood is close to that of a Newtonian fluid, and no particle ejection occurs during a blood droplet impact \cite{laan2014a,degoede2018}.
They conjectured that blood can be approximated as a Newtonian fluid because the droplet is subjected to a high shear rate ($>10^3$  \si{s^{-1}}) during an impact, and the the viscosity of blood is constant under high shear rates  (Fig. \ref{fig:vis}).

The validity and applicability of this assumption (that the blood behaves as a Newtonian fluid during impact) has been confirmed for the spreading and splashing of blood, but not for spreading/splashing of blood simulants, which are particles-containing solutions with viscosity similar to that of blood.
It is important to confirm the validity of the assumption on blood simulants to identify the suitable kind of particles to be used to mimic blood for the development of blood simulants for bloodstain pattern analysis.
Blood simulant is further discussed in this section in a later paragraph.
It should be noted that there are significant differences in deformability between red blood cells \cite{schmid-schonbein1969} and the resin or ceramic particles used by Grishaev {\it et al}. and Peters {\it et al}.
Previous research showed that a red blood cell, the Young's modulus of which is 1-10 kPa \cite{doshi2009,yeow2017}, has high deformability \cite{shin2005,liu2006,sasaki2018,nader2019}.
The deformability of red blood cells possibly plays a role in blood droplet impact.
In addition, because plasma has elasticity due to its protein components \cite{brust2013,kolbasov2016}, whole blood is a viscoelastic fluid.
Although de Goede {\it et al}. \cite{degoede2021} reported that plasma elasticity is negligible in a blood droplet impact, their study investigated only the spreading of blood droplets, without investigating the effect of elasticity on the splashing.

\begin{figure}[t]
    \centering
    \includegraphics[width=1\columnwidth]{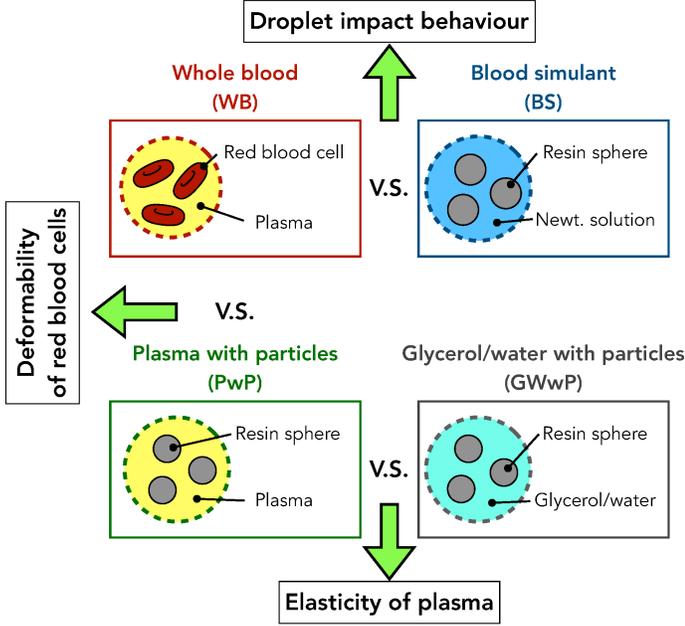}
    \caption{Comparison of droplet impact behaviour of four different solutions.}    \label{fig:method}
\end{figure}

In this study, we investigated the effects of the deformability of red blood cells and the elasticity of plasma on droplet impact to understand the Newtonian behaviour of blood by conducting comparative droplet impact experiments using four different solutions (Fig. \ref{fig:method}).To consider the effect of the deformability of red blood cells, we compared the droplet impact behaviour of whole blood (WB), in which red blood cells are dispersed in plasma, and a solution of plasma with particles (PwP), in which resin particles are dispersed in plasma.
In the experiment conducted to consider the effect of plasma elasticity, we compared the droplet impact behaviour of PwP and glycerol/water with particles (GWwP), in which resin particles are dispersed in a glycerol/water solution.

As previously mentioned, understanding the role of blood components in droplet impact behaviour is important for the development of blood simulants.
Studies on bloodstain pattern analysis and blood droplet impact are generally based on human or animal blood \cite{laan2014a,degoede2018,raymond1996}.
Such studies raise ethical and physiological concerns.
To address such concerns, it is necessary to develop a blood simulant (BS) made of non-biological materials.
Brutin {\it et al}. have demonstrated that red blood cells will affect the dried bloodstain pattern \cite{bruin2011}.
Besides, Smith et al. found a correlation between dried bloodstain pattern and droplet impact energy \cite{smith2018b}.
These studies indicate that particles that simulate red blood cells need to be added to the BS used for bloodstain pattern analysis.
Currently, there are several types of BSs \cite{stotesbury2017a,lee2020}.
However, their impact behaviours have not been fully investigated.
Thus, we also investigated the droplet impact behaviour of a commercially available BS, that contains resin particles to simulates the viscosity of WB.
The BS was then compared with WB to explore issues related to the development of blood simulants for bloodstain pattern analysis (Fig. \ref{fig:method}).

\section{Experimental method}

This section describes the preparation of the four solutions used in this study (Sec. \ref{sec:solutions}) and the droplet impact experiment (Sec. \ref{sec:experiment}).

\subsection{Solutions}\label{sec:solutions}

We obtained WB from a grown beagle dog because dog blood is quite similar to human blood in terms of the size, shape and volume fraction of red blood cells, and physical properties, which are namely the surface tension, viscosity, and density \cite{huggins1971,rebar1981}.
Raymond {\it et al}. showed that the droplet impact behaviour of the animal blood, which has similar physical properties to those of human blood, is quite similar to that of human blood \cite{raymond1996}.
In addition, the particle size is also an important parameter in the droplet impact of suspensions \cite{grishaev2015,peters2013a,grishaev2019}.
Therefore, this study uses dog's blood as a substitute for human blood in the droplet impact behaviour.
The volume fraction of red blood cells in blood is called haematocrit, and it is approximately 45 vol\% in human blood \cite{sousa2016}.
The human red blood cells have a biconcave shape with a diameter of approximately 8 \si{\um} \cite{sousa2016}.
The dog's red blood cells have the same shape as human's \cite{rebar1981} and the its haematocrit is about 44.3 vol\% \cite{huggins1971}.
Blood contains white blood cells and platelets at less than 1 vol\% \cite{baskurt2003,huggins1971}.
We assumed that their effects on droplet impact are negligible because of their minute volume fraction.

For preparing PwP, plasma was obtained from WB by centrifuging.
Hydrophilic spherical resin particles (Negami Chemical Industrial Co. Ltd., Art pearl SE-006T) with a diameter of approximately 6 \si{\um} were then added to plasma.
The particles were made of acrylic and had a specific gravity of 1.2.
The Young's modulus of acrylic, the material of resin particles, is about 2-6 GPa \cite{ashcroft1996,ishiyama2002}, which is more than $10^6$ times higher than that of red blood cells.
To make the viscosity of PwP similar to that of WB, the volume fraction of particles was adjusted to 25 vol\%.

For GWwP, the same 6 \si{\um} spherical particles (Negami Chemical Industrial Co. Ltd., SE-006T) were dispersed in a glycerol/water solution at 25 vol\% to make the viscosity similar to that of WB.
The glycerol/water solution was a mixture of 20 wt \% glycerol (Wako Pure Chemical Industries) in water.
The glycerol/water solution does not have elasticity, in contrast with plasma.

The commercially available BS was purchased from Yamashina Seiki Co., Ltd.
The product code is PB-10W.
BS consists of white spherical resin beads with a diameter of about 10 \si{\um} dispersed (45 vol\%) in a water-based Newtonian solution to mimic red blood cells.
The specific gravity of the resin beads was 1.2.

\begin{figure}[t]
    \centering
    \includegraphics[width=1\columnwidth]{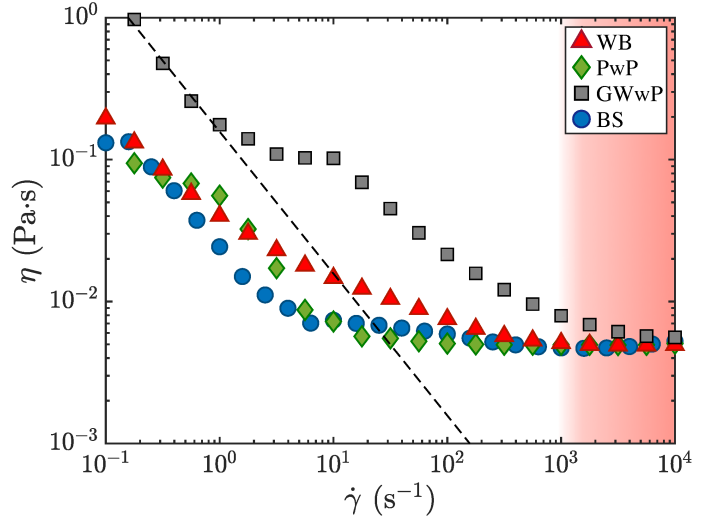}
    \caption{Dependence of shear viscosity $\eta$ upon the shear rate $\dot \gamma$ for the solutions used this study. The dashed line represents the minimum shear viscosity measurable by the rheometer.
    The range of high shear rate to which the impacting droplets are subjected is highlighted in red.}
    \label{fig:vis}
\end{figure}

The viscosity of each solution measured by a shear rheometer (TA Instruments Inc., AR-G2) is shown in Fig. \ref{fig:vis}.
All solutions show non-Newtonian characteristics, with the viscosity $\eta$ decreasing with increasing shear rate $\dot \gamma$ (shear-thinning).
The viscosity of all solutions approaches about 5 $\rm m Pa\cdot s$ at a high shear rate ($>10^3$ \si{s^{-1}}).
Because the droplets are subjected to a high shear rate ($>10^3$ \si{s^{-1}}) during impact, the viscosity is adopted as the effective viscosity of each solution in this study.
Table \ref{tab:pp} shows the density $\rho$, the surface tension $\sigma$, and the viscosity $\eta$ for each solution at the high shear rate $\dot \gamma = 10^4$ \si{s^{-1}}.
The surface tension was measured by the pendant-drop method \cite{berry2015}.
The density was measured by a pycnometer (AS ONE Co., Specific Gravity Bottle 5 mL).

\begin{table*}[t]
    \centering
    \caption{Physical properties of the solutions used in this study.}
    \small
    \begin{tabular*}{\textwidth}{@{\extracolsep{\fill}}lccc}
\hline
Solution (abbreviation)                           & Density $\rho$ (\si{kg/m^3}) & Surface tension $\sigma$ (\si{mN/m}) & Viscosity $\eta$ (\si{m \pascal \cdot s}) \\ \hline
Whole blood (WB)                    & 1058.4                                                                         & 67.7                                                                     & 4.9                                                                                             \\
Plasma with particles (PwP)          & 1068.7                                                                         & 43.1                                                                     & 5.0                                                                                             \\
Glycerol and water with particles (GWwP) & 1098.2                                                                         & 41.0                                                                      & 5.6                                                                                             \\
Blood simulant (BS)                 & 1059.5                                                                         & 71.5                                                                      & 5.0                                                                                             \\
Human's blood \cite{hulse-smith2005,lee2020,raymond1996,decastro2016,laan2014,rosina2007}                 & 1050-1063                                                                         & 47.7-63.1                                                                     & 3.8-5.6                                                                                             \\
\hline
\end{tabular*}
    \begin{flushleft}
    \end{flushleft}
    \label{tab:pp}
\end{table*}

\subsection{Droplet impact experiment}\label{sec:experiment}

\begin{figure}[t]
    \centering
    \includegraphics[width=0.85\columnwidth]{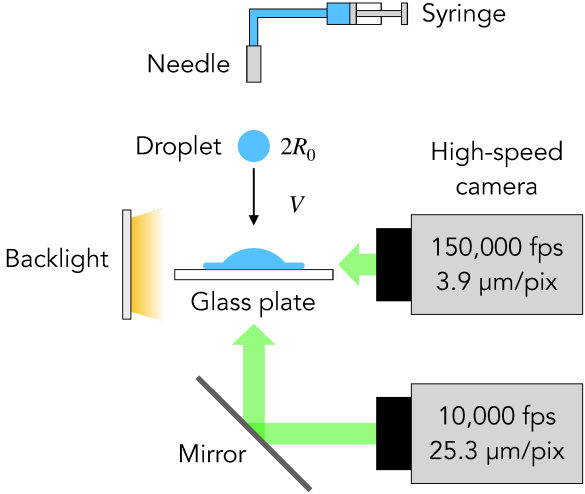}
    \caption{Schematic of the experimental setup with all the major components labelled.}
    \label{fig:setup}
\end{figure}

A schematic of the experimental setup is shown in Fig. \ref{fig:setup}.
A droplet was formed at the tip of a needle (Musashi Engineering Inc., PN-18G-A, 1.27-mm outer diameter) connected to a syringe.
The droplet detached from the needle by its weight, fell freely, and then impacted a hydrophilic glass plate (Muto Pure Chemicals Co., Ltd., Starfrost 5116) placed under the needle.
The impact behaviour of droplets was captured from the side and bottom simultaneously by two high-speed cameras (Fig. \ref{fig:setup}).
From the side, a high-speed camera (Photron Ltd., FASTCAM SA-Z) with a temporal resolution of 150,000 fps and a spatial resolution of 3.9 \si{\um}/pix was used to visualise the fast and small secondary droplets.
From the bottom, a high-speed camera (Photron Ltd., FASTCAM SA-X) with a temporal resolution of 10,000 fps and a spatial resolution of 25.3 \si{\um}/pix was used to visualise the spreading behaviour, the temporal evolution of the spreading radius $R(t)$, and the maximum radius $R_{max}$, via a mirror placed under the glass plate.
The impact velocity $V$ was varied in the range 2.0-4.5 \si{m/s} by adjusting the height of the needle over the glass plate between from 60 to 160 cm.
Three to four measurements were conducted at each height.
$V$ was calculated from the position of the bottom of the droplet at 10-12 frames before the droplet impacted the surface.
The maximum relative uncertainty of $V$ is about 5 \%.
The initial radii $R_0$ of the WB, PwP, GWwP, and BS droplets were 1.64$\pm$0.08, 1.46$\pm$0.05, 1.52$\pm$0.04, and 1.74$\pm$0.02 \si{mm}, respectively.
All experiments were performed for $Oh > 0.0136$, where no prompt splashing is expected to occur \cite{palacios2013}, and were carried out at atmospheric pressure and room temperature (about 25 $^\circ$C).

\section{Results}\label{sec:results}

This section reports the results obtained from the experiment.
The morphology during the impact, which is important in bloodstain pattern analysis, is shown in Sec. \ref{sec:morphology}.
The spreading factor is reported in Sec. \ref{sec:spread_f} and measurement of the splashing threshold is described in Sec. \ref{sec:splash_t}.

\subsection{Impact morphology}\label{sec:morphology}

We consider the morphology of each solution in terms of the Weber number, which is the ratio of inertial force to surface tension ($We = \rho R_0 V^2 / \sigma$) \cite{mehdizadeh2004}.

\begin{figure*}
    \centering
    \includegraphics[width=0.82\textwidth]{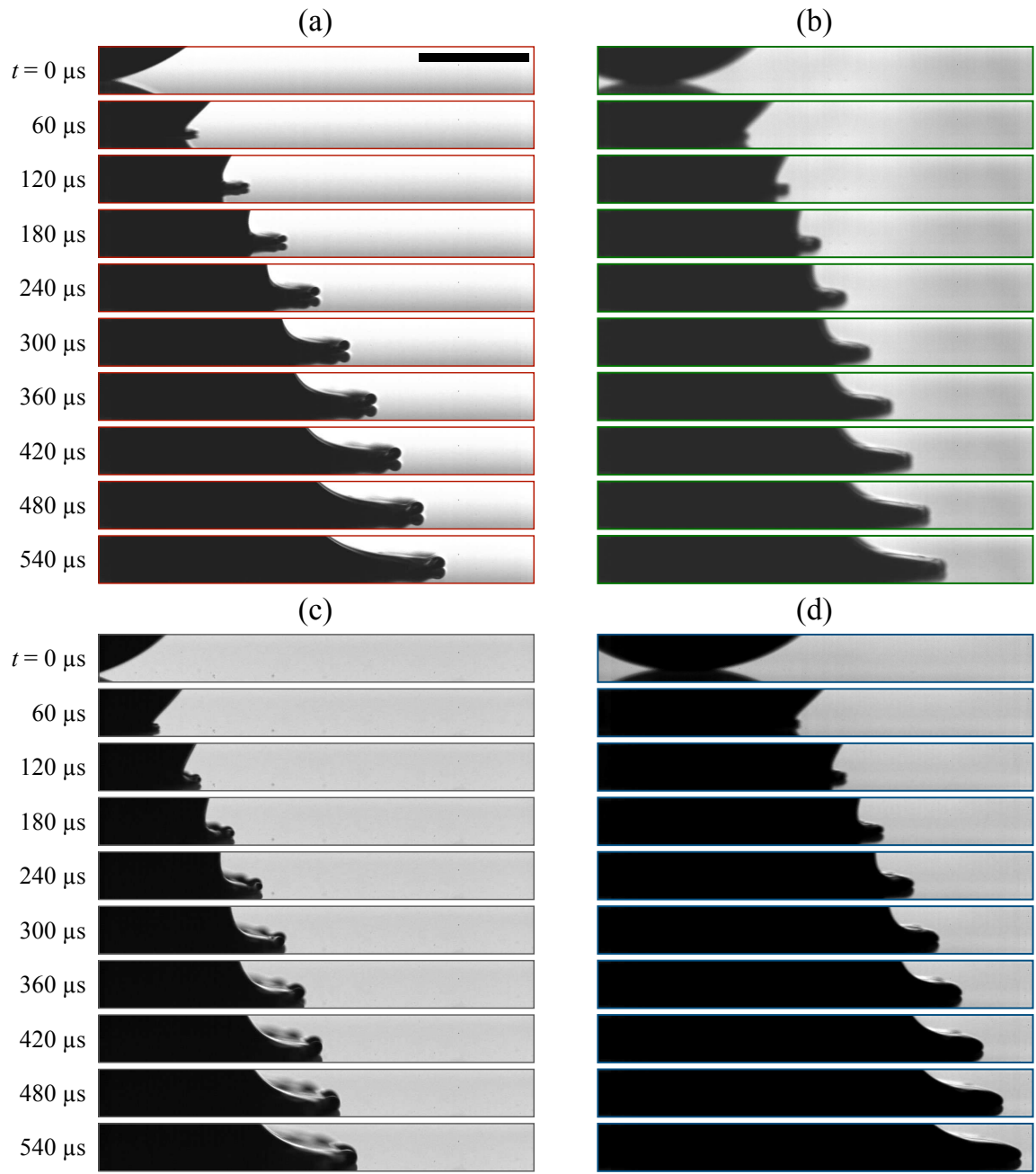}
    \caption{side-veiw image of impacting droplets at $We = 200 \pm 25$ for (a) whole blood (WB) ($We = 182$, $Re = 948$), (b) plasma with particles (PwP) ($We = 206$, $Re = 755$), (c) glycerol/water with particles (GWwP) ($We = 217$, $Re = 864$), and (d) blood simulant (BS) ($We = 186$, $Re = 991$). The scale bar [solid black line in (a)] corresponds to 1 \si{mm}.
    }
    \label{fig:side_spread}
\end{figure*}

\begin{figure*}
    \centering
    \includegraphics[width=0.87\textwidth]{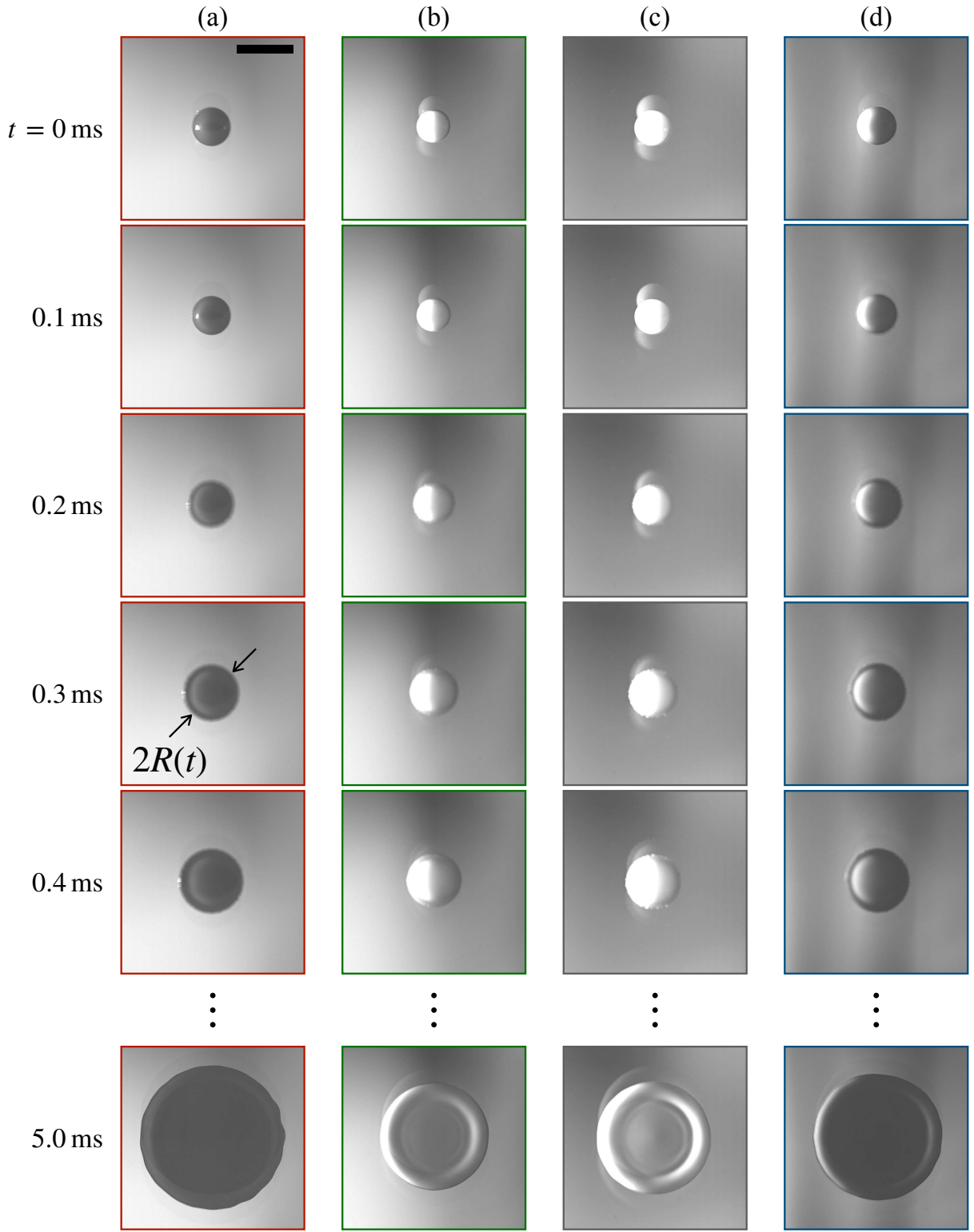}
    \caption{bottom-veiw image of impacting droplets at $We = 200 \pm 25$ of (a) WB ($We = 182$, $Re = 948$), (b) PwP ($We = 206$, $Re = 755$), (c) GWwP ($We = 217$, $Re = 864$), and (d) BS ($We = 186$, $Re = 991$). The scale bar [solid black line in (a)] corresponds to 5 \si{mm}].
    }
    \label{fig:bottom_spread}
\end{figure*}

For $We = 200 \pm 25$, the image sequence taken from the side is displayed in Fig. \ref{fig:side_spread}.
We defined $t=0$ as the time at which the droplet contacts the solid surface.
Fig. \ref{fig:side_spread} shows that the droplets of all solutions spread without splashing.
The image sequence taken from the bottom displayed in Fig. \ref{fig:bottom_spread} also shows that all solutions exhibit no splashing.
The spreading radius $R(t)$ at $t = 5.0$ ms (approximately equal to the maximum radius $R_{max}$) is the largest for WB, followed by BS, GWwP, and PwP.
Note that the Reynolds number, which is the ratio of the inertial to viscous forces ($Re=\rho R_0V/\eta$) is the highest for WB, followed by BS, GWwP, and PwP.
The maximum radius $R_{max}$ of each solution is discussed in detail in the next section.

\begin{figure*}
    \centering
    \includegraphics[width=0.82\textwidth]{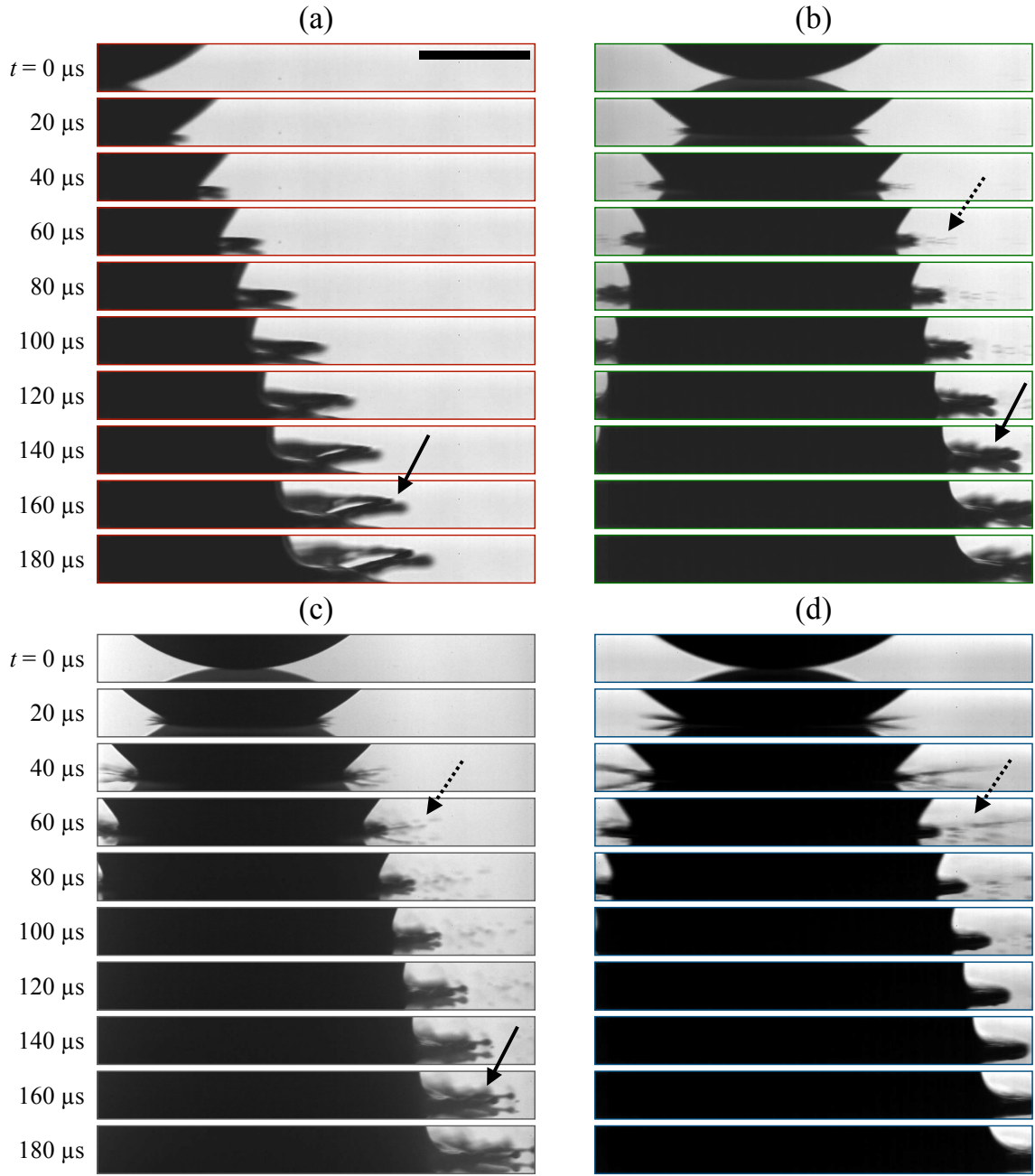}
    \caption{side-veiw image of impacting droplets at $We = 550 \pm 25$ of (a) WB ($We = 526$, $Re = 1634$), (b) PwP ($We = 554$, $Re = 1236$), (c) GWwP ($We = 569$, $Re = 1115$), and (d) BS ($We = 539$, $Re = 1697$). The solid and dashed arrows point to finger- and tiny-splashing, respectively (Fig. \ref{fig:splash_type}). The scale bar [solid black line in (a)] corresponds to 1 \si{mm}.
    }
    \label{fig:side_splash}
\end{figure*}

The side-veiw images taken for $We = 550 \pm 25$ are shown in Fig. \ref{fig:side_splash}.
As indicated by the dashed arrows, secondary droplets with a diameter of about 10 \si{\um} were ejected from the tip of the lamella for PwP, GWwP, and BS, but not WB.
This splashing resembles the prompt splashing mentioned in Sec. \ref{sec:intro}, despite the droplet impact experiments being performed at $Oh>0.0062$, where no prompt splashing is expected to occur \cite{palacios2013}.
As indicated by the solid arrows in Fig. \ref{fig:side_splash}, for WB, PwP, and GWwP, the lamella lifted off from the glass plate and formed a "finger" structure, and then disintegrated into secondary droplets with diameters of a few hundred micrometres.
This type of splashing is similar to the corona splashing described in Sec. \ref{sec:intro}.
This splashing did not occur in the case of BS.

\begin{figure*}
    \centering
    \includegraphics[width=1\textwidth]{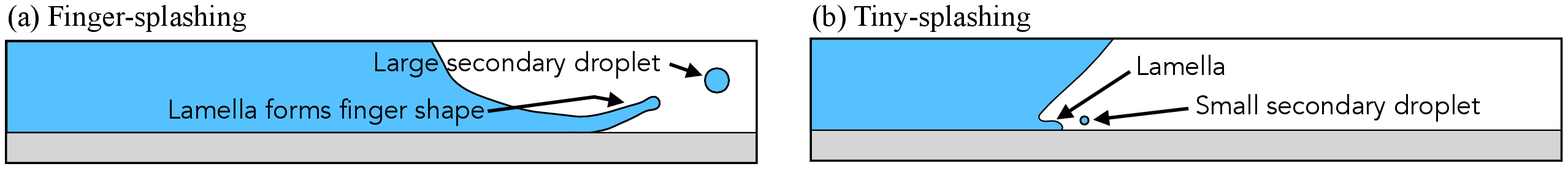}
    \caption{Illustration of the splashing morphology. (a) Finger-splashing: the lamella forms a "finger" structure and produces large secondary droplets. (b) Tiny-splashing: small secondary droplets are ejected from the tip of the lamella.}
    \label{fig:splash_type}
\end{figure*}

\begin{figure*}
    \centering
    \includegraphics[width=0.87\textwidth]{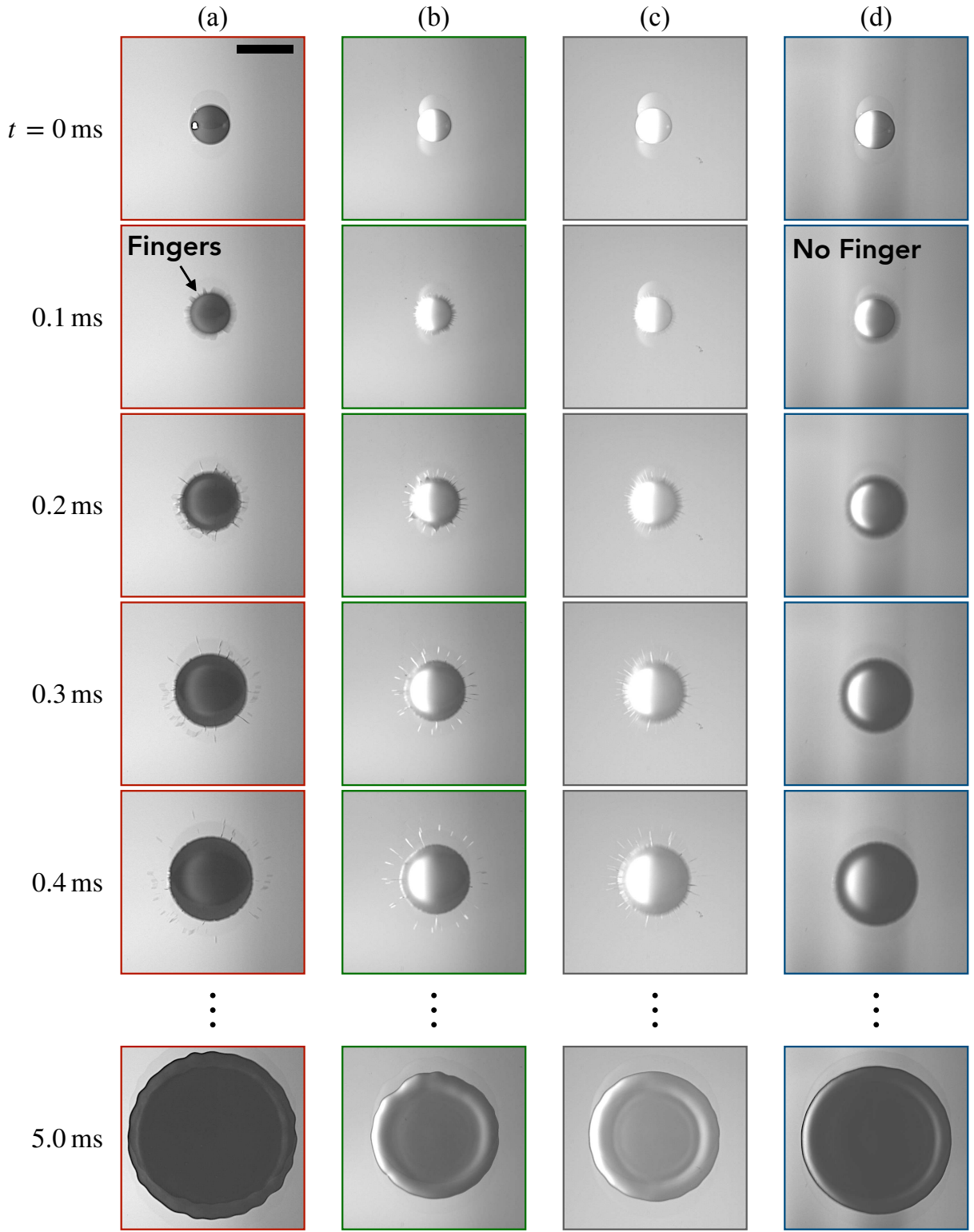}
    \caption{bottom-veiw image of impacting droplets at $We = 550 \pm 25$ of (a) WB ($We = 526$, $Re = 1634$), (b) PwP ($We = 554$, $Re = 1236$), (c) GWwP ($We = 569$, $Re = 1115$), and (d) BS ($We = 539$, $Re = $1697). The scale bar [solid black line in (a)] corresponds to 5 \si{mm}.
    }
    \label{fig:bottom_splash}
\end{figure*}

Because two types of splashing occur in the region of $Oh$ where only corona splashing is expected to occur, we named the splashing that formed fingers ``finger-splashing'' (Fig. \ref{fig:splash_type}(a)), and the splashing that ejected small secondary droplets ``tiny-splashing'' (Fig. \ref{fig:splash_type}(b)).
The image sequence taken for $We = 550 \pm 25$ from the bottom shows that WB, PwP, and GWwP exhibited finger-splashing [see Fig. \ref{fig:bottom_splash}(a), 0.1 ms].
In contrast, BS formed no fingers  (Fig. \ref{fig:bottom_splash}(d)).
Note that the tiny-splashing is invisible from the bottom-veiw image because the size of the secondary droplets is smaller than the spatial resolution (25.3 \si{\um}/pix).

In summary, WB exhibited only finger-splashing, while PwP and GWwP exhibited both finger-splashing and tiny-splashing.
BS exhibited only tiny-splashing.

\subsection{Spreading factor}\label{sec:spread_f}

\begin{figure}[t]
    \centering
    \includegraphics[width=1\columnwidth]{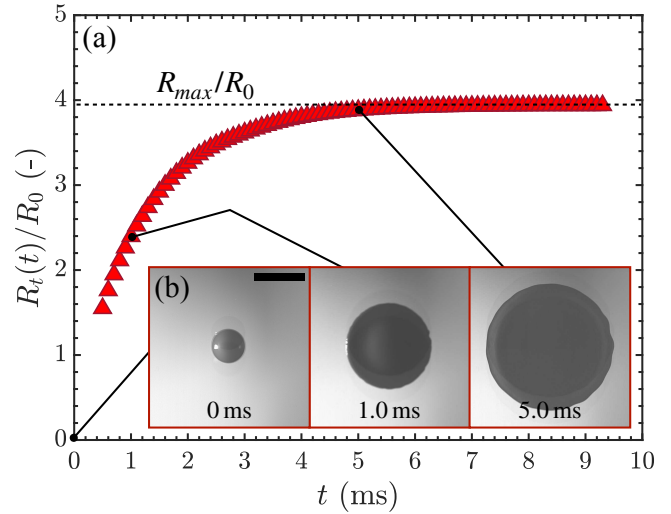}
    \caption{(a) Time evolution of spreading radius $R(t)$ of WB ($We = 182$, $Re = 948$). $t=0$ is defined as the time at which the droplet contacts the solid surface. The spreading radius $R(t)$ eventually reaches its maximum radius $R_{max}$. (b) bottom-veiw image of the impacting droplet at different times. The scale bar [solid black line shown in (b)] corresponds to 5 \si{mm}.}
    \label{fig:spreading}
\end{figure}

Figure \ref{fig:spreading} shows the spreading radius of WB at $We \simeq 182$.
After impact, the spreading radius $R(t)$ of the droplet increased and eventually approached its maximum radius $R_{max}$.

The spreading factor $R_{max}/R_0$ is a crucial parameter for reconstructing a crime scene through bloodstain pattern analysis \cite{laan2014a,attinger2013}.
Many models of the spreading factor have been proposed \cite{wang2019}.
Pasandideh-Fard \textit{et al}. developed a model based on energy dissipation due to the liquid viscosity \cite{pasandideh-fard1996}.
They suggested that when the inertia of an impacting droplet is sufficiently high ($We \gg Re^{1/2}$), the spreading factor is a function of the Reynolds number:

\begin{equation}
    \frac{R_{max}}{R_0} = C Re^{1/4},
    \label{eq:sf}
\end{equation}

\noindent where $C$ is the fitting constant.
The scaling law in Eq. (\ref{eq:sf}) is commonly used in the field of bloodstain pattern analysis because many experiments on blood droplet impact \cite{adam2012,hulse-smith2005,attinger2013}, including ours, satisfy the condition $We \gg Re^{1/2}$.

\begin{figure}[t]
    \centering
    \includegraphics[width=1\columnwidth]{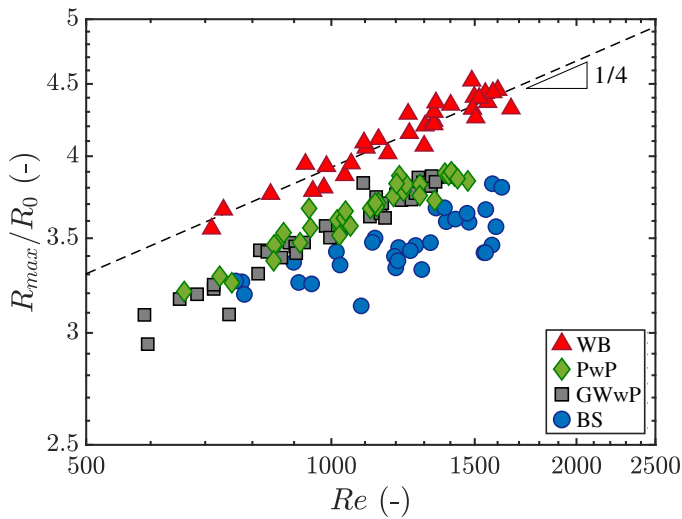}
    \caption{Spreading factor for each solution as a function of Reynolds number. The dashed line corresponds to $R_{max}/R_0 \propto Re^{1/4}$.}
    \label{fig:spread_f}
\end{figure}

Fig. \ref{fig:spread_f} shows the spreading factor as a function of $Re$.
Remarkably, although the shear viscosity of the solutions (PwP, GWwP, and BS) is similar to that of WB (see Fig. \ref{fig:vis}), their spreading factors are smaller than that of WB.
The spreading factor $R_{max}/R_0$ of WB increases in proportion to $Re^{1/4}$.
The spreading factors for PwP and GWwP are smaller than that for WB, although they also increase in proportion to $Re^{1/4}$.
The spreading factor of BS is the smallest in comparison with the other three solutions.
Besides, it is more scattered than the others, most likely due to to the difference in the local viscosity, i.e. the local volume fraction of particles, in the droplet of BS.
Although we tried to minimise the variation of local volume fraction of particles, BS with a high volume fraction of particles tend to aggregate more easily than those in PwP and GWwP.

Note that, the shear viscosity of GWwP for high shear rate is approximately 0.6 mPa$\cdot$s greater than that of other solutions (Table. \ref{tab:pp}).
We discuss how much the difference in shear viscosity affects the spreading factor.
The relationship between $\phi$ and $\eta$ of suspension can be described by Krieger-Dougherty model \cite{krieger1959} which is expressed in the equation $\eta = \eta_s \left( 1 - \phi/\phi_{m} \right)^{-n},$ where $\eta_s$ is the viscosity of the solvent and $\phi_{m}$ and $n$ are fitting parameters.
By fitting this equation to our measurement results of glycerol/water with particles, we obtain $\phi_m$ = 0.62 and $n$ = 2.42.
Using these values, the volme fraction of particles $\phi$ should be 0.24 (24 vol\%) for GWwP to obtain the same viscosity of WB (5 mPa$\cdot$s).
Therefore, the difference in $\phi$ is $\Delta\phi \sim$ -0.01 (-1 vol\%).
The relationship between the change in spreading factor ($R_{max}/R_0$) and the change in volume fraction of particles is reported to be $\Delta(R_{max}/R_0)/\Delta\phi \sim$ -5.6 \cite{grishaev2015}.
Thus the change in $R_{max}/R_0$ is approximately +0.056 when the concentration of GWwP changes by approximately -0.01 (-1 vol\%).
This change in $R_{max}/R_0$ is significantly smaller than the difference between $R_{max}/R_0$ for GWwP and other solutions.
Therefore it does not significantly affect the results for spreading.

\subsection{Splashing threshold}\label{sec:splash_t}

\begin{figure}[t]
    \centering
    \includegraphics[width=1\columnwidth]{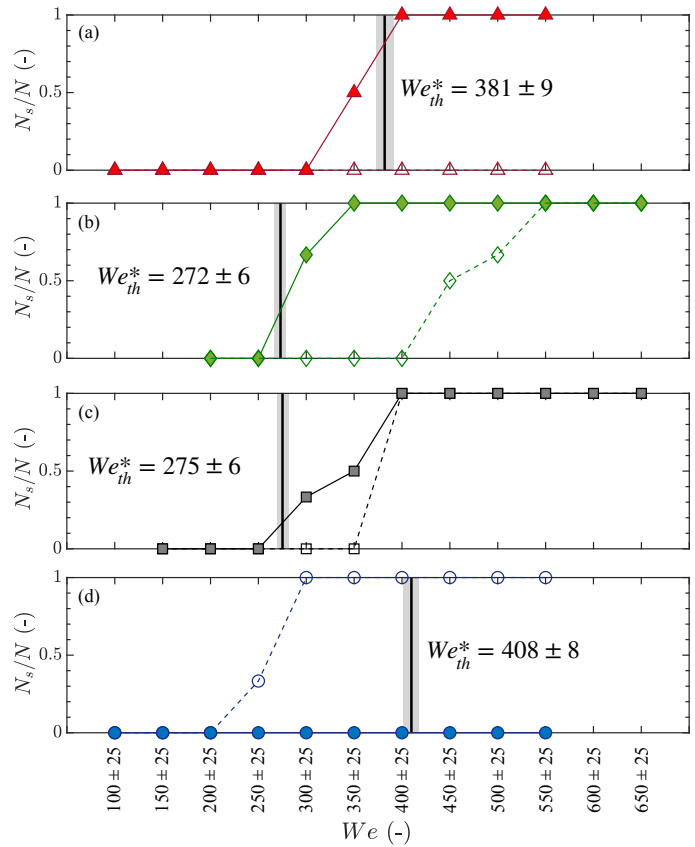}
    \caption{Probability of splashing occurrence $N_s/N$ of each $We$ range compared with the splashing threshold $We_{th}^*$ predicted by de Goede's model  (Eq. (\ref{eq:st})) \cite{degoede2018} with $\beta = 0.112\pm0.001$ for (a) WB, (b) PwP, (c) GWwP, and (d) BS. Closed symbols indicate finger-splashing, and open symbols indicate tiny-splashing.
    } 
    \label{fig:splash_t}
\end{figure}

To find the splashing threshold, Fig. \ref{fig:splash_t} shows the probability of splashing occurrence $N_s/N$ for each $We$ range, where $N$ is the number of repetitions of the experiment for the respective Weber number range and $N_s$ is the number of observed splashing.
The Weber number at which $N_s/N$ is greater than 0 is regarded as the splashing threshold $We^\ast$.
Splashing threshold for each solutions are shown in Table \ref{tab:sp_th}.

To confirm that WB can be approximated as a Newtonian fluid during splashing as well as spreading, the splashing threshold for WB was compared with the theoretical prediction for Newtonian fluids.
A model to predict the threshold for corona splashing (i.e., finger-splashing) for a Newtonian fluid was proposed by Riboux and Gordillo \cite{riboux2014,riboux2017}.
de Goede {\it et al}. \cite{degoede2018} simplified the model and showed that the splashing threshold for blood can be predicted by the following equation \cite{degoede2018}:
\begin{equation}
    V_{th}^\ast = \left( \frac{\sigma^{2/3}\beta \tan(\alpha)}{2.22 \sqrt{\eta_g} (\rho R_0)^{1/6} } \right)^{6/5},
    \label{eq:st}
\end{equation}
where $\eta_g$ is the viscosity of the surrounding air ($\sim 1.85 \times 10^{-5}$ Pa$\cdot$s \cite{riboux2014}); $\alpha$ is the wedge angle, which is the angle between the lifted lamella and the solid surface; and $\beta$ is the splashing ratio.
Riboux and Gordillo show that the model agrees well with experimental results when the wedge angle is set to 60 $^\circ$ \cite{riboux2014}, which is experimentally confirmed by de Goede {\it et al}. \cite{degoede2018}.
Therefore, we also used $\alpha = 60 \, ^\circ$.
de Goede {\it et al}. proposed that $\beta$ for a Newtonian fluid is $0.120 \pm 0.008$ and $\beta$ for blood is $0.112\pm0.001$ \cite{degoede2018}.

\begin{figure}[t]
    \centering
    \includegraphics[width=0.95\columnwidth]{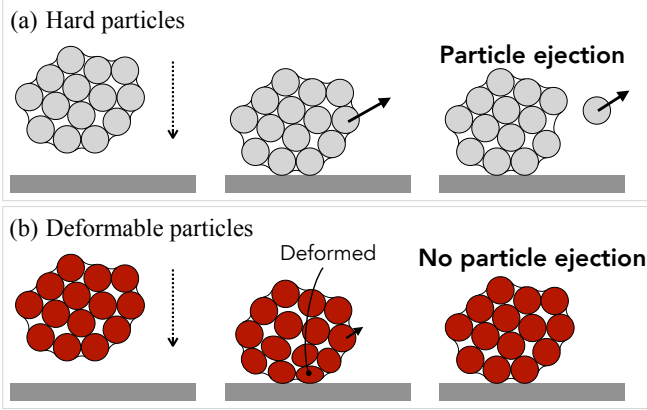}
    \caption{Mechanism of particle ejection upon droplet impact for (a) hard particles and (b) deformable particles.}
    \label{fig:pe}
\end{figure}

While WB shows only finger-splashing, PwP shows both finger-splashing and tiny-splashing  (Fig. \ref{fig:side_splash}(a,b,c)).
In the case of PwP, only finger-splashing occurs, while no tiny-splashing occurs in a certain range of $We$ (Fig. \ref{fig:side_splash}(b)).
Additionally, the size of the secondary droplets ejected by tiny-splashing is several dozen micrometres, which is similar to the diameter of the particles dispersed in the respective solutions.
These results suggest that tiny-splashing is not hydrodynamic splashing but actually ``particle ejection,'' in which particles dispersed in the solution are ejected.
Such particle ejection is a typical splashing behaviour of suspensions \cite{peters2013a,grishaev2017}: the particles near the droplet interface are ejected as a result of the transfer of the droplet momentum among neighbouring particles during the impact  (Fig. \ref{fig:pe}(a)).
A previous study suggested that particle ejection depends on particle properties such as hardness \cite{peters2013a}.
Therefore, the reason why WB, a suspension of red blood cells, did not exhibit particle ejection may be related to the high deformability of red blood cells.
The details are discussed in Sec. \ref{sec:effect of rbc}.

The splashing threshold $We_{th}^\ast$ predicted by the theoretical model \cite{degoede2018} for each solution is shown in Fig. \ref{fig:splash_t}.
$We_{th}^\ast$ for WB is in a good agreement with the measured value $We^\ast$.
This suggests that blood behaves as a Newtonian fluid during splashing.
For both PwP and GWwP, the model can predict the measured threshold for finger-splashing.
Conversely, BS does not show finger-splashing, which is against the prediction of the model.

The difference of viscosity between GWwP and other solutions ($\Delta \eta \sim$ -0.6 mPa$\cdot$s) produces the change in the volume fraction of particles $\Delta \phi \sim$ -0.01 (-1 vol\%) (see Sec. \ref{sec:spread_f}).
Here we also estimate how much this difference affects a splashing threshold by considering the results of another study by Grishaev {\it et al}. \cite{grishaev2019}.
They suggested the splashing criterion of suspensions as a function of the volume fraction of particles with a particle size of approximately 10 \si{\um}, which is similar to our case.
According to their splash criterion, decreasing volme fraction of particles by $\Delta \phi \sim $ -0.01 (-1 vol\%) leads to an increase in the splashing threshold $We^\ast$ of approximately 10.
Therefore, the results for splashing should not be significantly different either.

\begin{table*}[t]
    \caption{Splashing threshold We* of each solution: measured values and theoretical value predicted by Eq. (\ref{eq:st}).}
    \small
    \begin{tabular*}{\textwidth}{@{\extracolsep{\fill}}lccc}
    \hline
    Solution & $We^\ast$ for tiny-splashing & $We^\ast$ for finger-splashing & $We^\ast_{th}$ predicted by Eq. (\ref{eq:st}) \\ \hline
    WB       & -                            & 350$\pm$25                     & 381$\pm$9                      \\
    PwP      & 450$\pm$25                   & 300$\pm$25                     & 272$\pm$6                      \\
    GWwP     & 400$\pm$25                   & 300$\pm$25                     & 275$\pm$6                      \\
    BS       & 250$\pm$25                   & -                              & 408$\pm$8                      \\ \hline
    \end{tabular*}
    \label{tab:sp_th}
\end{table*}

\section{Discussion}

To consider the effect of the deformability of red blood cells (Sec. \ref{sec:effect of rbc}) and the elasticity of plasma (Sec. \ref{sec:effect of plasma}), we discuss the results for WB, PwP, and GWwP (Fig. \ref{fig:method}).
Furthermore, to explore the challenges in developing BSs for bloodstain pattern analysis, we compare the results for WB and BS (Sec. \ref{sec:discuss BS}).

\subsection{Effect of the deformability of red blood cells}\label{sec:effect of rbc}

Both WB and PwP exhibited splashing, which forms fingers.
Their spreading morphology was similar.
However, the spreading factor for PwP is smaller than that for WB (Fig. \ref{fig:spread_f}).
This difference is mainly due to differences in the particles dispersed within them, namely, red blood cells and resin particles.
In the case of a suspension with hard particles, such as PwP, Grishaev {\it et al}. \cite{grishaev2015} suggested that the interaction of a particle with other particles and with the solid surface (such as energy dissipation due to friction) reduces the spreading factor.
Based on this suggestion, the high spreading factor for WB indicates that such energy dissipation is unlikely to occur due to the high deformability of red blood cells.
Note that the high deformability of red blood cells play a similar role in blood vessels, where the deformability reduces the frictional resistance between red blood cells and the vessel wall \cite{zhou2006,sasaki2018}.

PwP showed both particle ejection and finger-splashing, whereas WB showed only finger-splashing.
This implies that the high deformability of red blood cells suppresses the transfer of momentum between the particles, resulting in no particle ejection  (Fig. \ref{fig:pe}(b)).
Therefore, the aforementioned effects of the deformability of red blood cells may be the reason that leads to the splashing threshold for WB being able to be predicted by the theoretical model for Newtonian fluids \cite{degoede2018}. 

\subsection{Effect of plasma elasticity}\label{sec:effect of plasma}

To discuss the effect of the elasticity of plasma, the results for GWwP and PwP are compared.
The plasma has elasticity, whereas the glycerol/water solution does not.

The spreading factors for PwP and GWwP are similar.
This result suggests that the effect of plasma elasticity on the spreading factor is negligible, as suggested by a previous study \cite{degoede2021}.

The threshold for tiny-splashing in PwP is higher than that in GWwP.
Vega and Castrej\'{o}n-Pita-Pita \cite{vega2017} performed droplet impact experiments using polymeric liquids, and reported that the elasticity of a liquid increased the threshold for tiny-splashing.
Therefore, the higher tiny-splashing threshold for PwP than for GWwP is likely due to the elasticity of the plasma.
In contrast, because there is no significant difference in threshold for finger-splashing between PwP and GWwP (Fig. \ref{fig:side_splash}), finger-splashing may not be affected by the elasticity of the plasma.

The volume fraction of particles in the studies by Grishaev {\it et al}. \cite{grishaev2015} and Peters {\it et al}. \cite{peters2013a} are in the range of 0.01-0.33 and 0.62, respectively.
These studies show that the volume fraction of particles affects the spreading and splashing behaviour of droplet impact.
Since volume fraction of particles in WB is approximately twice as large as those in PwP or GWwP, direct comparison between them may not be fair.
Nevertheless, if we recall the similar shear-viscosity curves, our results imply that the plasma elasticity may be negligible in the splashing of WB which exhibits only finger-splashing.
Additionally, the threshold for finger-splashing of WB agrees with the prediction by the model of a Newtonian fluid, similar to the result of de Goede {\it et al}. \cite{degoede2018}.

\subsection{Droplet impact behaviour of the commercial blood simulant}\label{sec:discuss BS}

As shown in Sec. \ref{sec:results}, the impact behaviour of BS is very different from that of WB; for example, it does not show the finger-splashing that is observed in WB.
Our results indicate that a solution that contains hard particles, although it has a similar viscosity as WB, cannot simulate the droplet impact behaviour of WB.
Thus, we recommend that deformable particles be used in BS for bloodstain pattern analysis.

\begin{figure}[t]
    \centering
    \includegraphics[width=0.9\columnwidth]{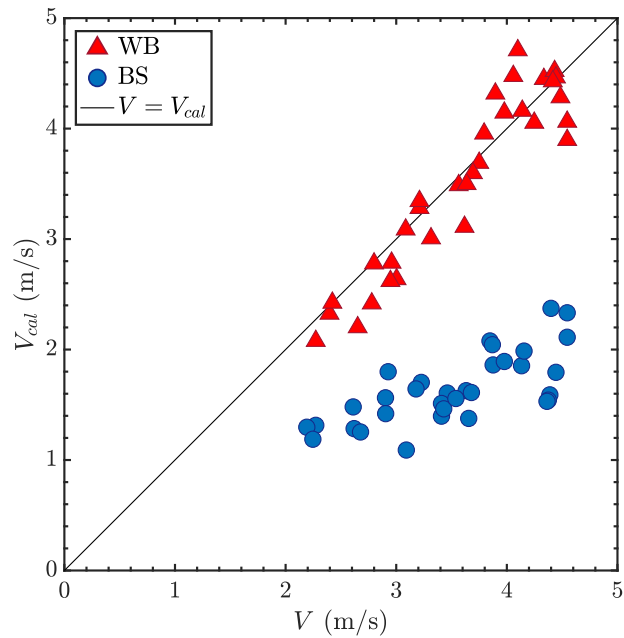}
    \caption{Impact velocity $V_{cal}$ estimated by Eq. (\ref{eq:sf}) correlates with experimentally measured impact velocity $V$.}
    \label{fig:V_Vcal}
\end{figure}

The difference in spreading factor between BS and WB leads to a significant error in the prediction of droplet impact velocity.
To quantify the difference between BS and WB in terms of the estimated impact velocity, we calculated the droplet impact velocity from the spreading factors for WB and BS as follows.
The proportionality constant $C$ in Eq. (\ref{eq:sf}) for WB is 0.70, which is obtained by fitting Eq. (\ref{eq:sf}) with WB's spreading factor data (Fig. \ref{fig:spread_f}).
By assuming that BS has the same proportional constant as WB, the estimated impact velocity for BS, $V_{cal}$, is shown as a black line in Fig. \ref{fig:V_Vcal}.
The estimated impact velocity for BS is much lower than that for WB and is underestimated by approximately 50 \%.

\section{Conclusions}

To elucidate why the droplet impact behaviour of blood can be approximated as a Newtonian fluid, we investigated the effect of the high deformability of red blood cells and the elasticity of plasma by conducting comparative experiments.
The effect of the deformability of red blood cells was evaluated by comparing WB and PwP, and the influence of the elasticity of plasma was evaluated by comparing PwP and GWwP.
We also compared WB with BS to explore the issues in the development of the BS used for bloodstain pattern analysis.
Each solution has non-Newtonian properties and a similar viscosity under a high shear rate ($>10^3$ \si{s^{-1}}).
The comparison of the impact behaviour was done using the spreading and splashing behaviour.

The spreading factor and splashing threshold for WB agree with those for Newtonian fluids, while those for PwP do not.
This difference is likely due to the high deformability of red blood cells.

It was found that the elasticity of the plasma did not affect the spreading behaviour, as also suggested by a previous study \cite{degoede2021}.
However, a slight difference in the threshold for particle ejection is found due to the elasticity of the plasma.

Previous studies have shown that blood can be approximated as a Newtonian fluid in terms of its spreading and splashing behaviour because the viscosity of blood is constant at high shear rates \cite{laan2014a,degoede2018}.
As for this study, the results suggest suggest that the high deformability of red blood cells is also essential for approximating blood as a Newtonian fluid during droplet impact.

A commercially available BS behaves quite differently from WB with regard to both spreading and splashing behaviour.
Thus, the addition of deformable particles to BS is highly recommended for bloodstain pattern analysis.

